\documentclass[12pt]{iopart}
\eqnobysec
\usepackage{amssymb}
\usepackage{xcolor}
\usepackage{graphicx}

\def\be{\begin{equation}}
\def\ee{\end{equation}}
\def\bea{\begin{eqnarray}}
\def\eea{\end{eqnarray}}
\def\ld*{{\cal L}_{d^*}}
\def\ld{{\cal L}_{d}}
\def\t0{t_0}

\begin{document}

\title[Mean number of encounters]{Mean number of encounters of $\bi{N}$ random walkers
and intersection of strongly anisotropic fractals}

\author{Lo\"\i c Turban}

\address{Groupe de Physique Statistique, D\'epartement Physique de la Mati\`ere et des Mat\'eriaux,
Institut Jean Lamour\footnote[1]{Laboratoire associ\'e au CNRS UMR 7198.}, CNRS---Universit\'e de Lorraine,\\
BP 70239, F-54506 Vand\oe uvre l\`es Nancy Cedex, France}
\ead{Loic.Turban@univ-lorraine.fr}

\begin{abstract} We study the mean number of encounters up to time $t$, $E_N(t)$, taking place in a subspace with dimension $d^*$ of a $d$-dimensional lattice, for $N$ independent random walkers starting simultaneously from the same origin. $E_N$ is first evaluated analytically in a continuum approximation and numerically through Monte Carlo simulations in one and two dimensions. Then we introduce the notion of the intersection of strongly anisotropic fractals and use it to calculate the long-time behaviour of $E_N$.
\end{abstract}


\maketitle

\section{Introduction}
Despite its long history, the study of random walks still remains an active field of research~\cite{spitzer64,weiss94,hughes95,redner01}. For the discrete random walk, introduced  by Polya \cite{polya21}, a quantity of interest for applications is the mean number 
$S_1(t)$ of distinct sites visited up to time $t$ by a single random walker. It evidently grows as $\sqrt{t}$ when $d=1$. The asymptotic behaviour for $d>1$ was obtained by Dvoretzky and Erd\"os \cite{dvoretzky51} with $S_1(t)\sim t/\ln t$ at the critical dimension $d_c=2$ and $S_1(t)\sim t$ for $d>2$. Exact values of the amplitudes on different lattices in $d=3$ and sub-dominant contributions were later calculated \cite{vineyard63,montroll65}.
Many other observables associated with the geometry of the support (i.e. the set of visited sites) of a random walk and their fluctuations have been studied more recently \cite{vanwijland97a,vanwijland97b}.

The peculiarity of the dimension $d=2$ was first noticed by Polya \cite{polya21} who showed that, with probability 1, a random walker returns infinitely often to the origin when $d\leq2$ and escape to infinity when $d>2$. This property is linked to the value of the fractal dimension of the random walk, $d_f=2$.

Instead of considering a single random walk, one may generalise to the case of $N$ statistically independent random walks starting from the same point. 
Such studies were first concerned with the properties of first passage times \cite{lindenberg80,weiss83,yuste96,yuste00b}.
The study of the number of distinct sites visited by $N$ random walkers, $S_N(t)$, was initiated in~\cite{larralde92} where asymptotic expressions for $N$ large were obtained. In~\cite{yuste99,yuste00a} some corrections to~\cite{larralde92} were given and sub-dominant contributions were evaluated.

The mean value of another interesting observable, $W_N(t)$, the mean number of common sites visited by $N$ random walkers up to time $t$, was studied in~\cite{majumdar12} where the asymptotic behaviour at long time was determined. Quite recently, exact expressions for the probability distributions of the number of distinct sites and the number of common sites visited by $N$ random walkers starting from the same origin have been obtained in $d=1$~\cite{kundu13}.

Following the work of Majumdar and Tamm~\cite{majumdar12}, we have shown how the asymptotic behaviour of $W_N(t)$ can be simply obtained exploiting the notion of fractal intersection~\cite{turban12}.
The rule governing the fractal dimension $d_f(N)$ of the intersection of $N$ statistically independent and isotropic fractals with dimensions $d_{f}$, embedded in Euclidean space with dimension $d$, is the following \cite{mandelbrot82}
\be 
d_f(N)=d-\overline{d_f(N)}\,,\qquad \overline{d_f(N)}=\min \left[d,N\overline{d_f}\right]\,,
\label{dfiso}
\ee
where $\overline{d_f}=d-d_f$ is the co-dimension \cite{codimension}. The dimension of the intersection vanishes at $d=d_c$, the upper critical dimension, when the sum of the co-dimensions is equal to $d$~\cite{dc}. 
The mean number of common sites visited, $W_N(t)$, behaves as the mass $M\sim R^{d_f(N)} $ of the fractal resulting from the intersection of the $N$ random walks. The typical size at time $t$ is $R\sim\sqrt{t}$ and the dimension of the intersection follows from~\eref{dfiso} with $d_f=2$.

In the present work we study the mean number of encounters up to time $t$, $E_N(t)$, of $N$ independent random walkers starting from a common origin at $t=0$. The events in space-time, corresponding to the meeting of the $N$ random walkers, take place at the intersection of the $N$ fractals associated with the support of the random walks~\cite{ew}. Since, when considered in space-time, a random walk is a strongly anisotropic fractal object, the scaling behaviour of $E_N(t)$ at long time is governed by the fractal dimension of the intersection of $N$ strongly anisotropic fractals.  

A fractal object is strongly anisotropic when its growth in different directions is governed by different exponents, $R_\parallel\sim R_\perp^z$.  The random walk in space-time, with a dynamical exponent $z=2$, is one of the simplest examples. Such a behaviour is not uncommon, it is observed in different contexts among which one may mention directed systems~\cite{privman89}, Lifshitz points~\cite{diehl05,henkel10} and some quantum critical points~\cite{sachdev99}, critical dynamics~\cite{hohenberg77} and non-equilibrium phase transitions~\cite{hinrichsen00,odor04,henkel08}.

The paper is organised as follows. In section 1, $E_N(t)$ is calculated analytically in a continuum approximation for encounters taking place in a subspace of dimension $d^*$ of the $d$-dimensional Euclidean space. 
In section 2, $E_N(T)$ is evaluated numerically through Monte Carlo simulations for discrete random walks in one and two dimensions. In section 3, the rule giving the dimension of the intersection of strongly anisotropic fractals is established, generalising~\eref{dfiso}. Then the notion of fractal intersection is used to evaluate the long-time behaviour of $E_N(t)$. In the final section, we formulate some remarks and present possible extensions.

\section{Mean number of encounters of $\bi{N}$ random walkers} 

We consider $N$ random walkers performing discrete, statistically independent random walks on a $d$--dimensional hyper-cubic lattice $\ld$ with lattice parameter $a$. The $N$ walks start simultaneously at $t=0$ from the origin of $\ld$. The position of a lattice site is given by $\vec{r}=\sum_{i=1}^d x_i\,\vec{u}_i$ where the $\vec{u}_i$ are mutually orthogonal unit vectors and $x_i=n_i\,a$ ($n_i\in\mathbb{Z}$). The walks are also discrete in time, $t=n\,\tau$. 

We want to calculate the mean cumulated 
number of encounters of the $N$ random walkers, occurring on a sub-lattice $\ld*$ 
of $\ld$, containing the origin, with dimension $d^*$~\cite{subspace} such that $0\leq d^*\leq d$. 

Let $I_N(\vec{r},t)$ be an indicator associated with a site at $\vec{r}$ for a given time $t$ with the following values:  $I_N(\vec{r},t)=1$ when the $N$  random walkers visit the site at $\vec{r}$ at the {\sl same time} $t$ and $I_N(\vec{r},t)=0$ otherwise. The number of encounters of the $N$ independent random walkers up to time $t$, for a realisation of the $N$ random walks, is given by $\sum_{t'=0}^t\sum_{\vec{r}\in\ld*}I_N(\vec{r},t')$. The mean cumulated number of encounters of the $N$ random walkers at time $t$, $E_N(t)$, is obtained by taking the average of this expression so that
\be 
E_N(t)=\sum_{t'=0}^t\sum_{\vec{r}\in\ld*}\langle I_N(\vec{r},t')\rangle=\sum_{t'=0}^t\sum_{\vec{r}\in\ld*}P_N(\vec{r},t')\,,
\label{ENt1}
\ee
where $P_N(\vec{r},t')$ is the probability to find the $N$ random walkers at $\vec{r}$ at time $t'$. 
For $N$ statistically independent random walks we have $P_N(\vec{r},t')=[P_1(\vec{r},t')]^N$. 

The $N$ random walkers starting simultaneously from the origin at $t=0$, their first encounter takes place at $t'=0$ and the first sum can be split as follows:
\be
E_N(t)=1+\sum_{t'=\tau}^t\sum_{\vec{r}\in\ld*}[P_1(\vec{r},t')]^N\,.
\label{ENt2}
\ee
Since we are interested in the long-time behaviour of $E_N(t)$ we use the continuum limit where both 
$\vec{r}$ and $t$ are treated as continuous variables. In this approximation
\be
P_1(\vec{r},t)\rightarrow\frac{\exp(-{r^2}/{4Dt})}{(4\pi Dt)^{d/2}}\,a^d\,,\quad \sum_{\vec{r}\in\ld*}\rightarrow \int\frac{d^{d^*}\!r}{a^{d^*}}\,,\quad \sum_{t'}\rightarrow\int\frac{d t'}{\tau}\,,
\label{P1}
\ee
where $D$ is the diffusion coefficient. Thus when $t\gg\tau$ one obtains
\bea
E_N(t)&\approx 1+\int_{\t0}^t\frac{d t'}{\tau}\int\frac{d^{d^*}\!r}{a^{d^*}}\frac{\exp(-{Nr^2}/{4Dt'})}{(4\pi Dt')^{Nd/2}}\,a^{Nd}\nonumber\\
&\approx 1+\frac{a^{Nd-d^*}}{\tau(4\pi D)^{Nd/2}}\int_{\t0}^t\frac{d t'}{(t')^{Nd/2}}A_N(t')\,,
\label{ENt3}
\eea
where $\t0=\tau/2$ will provide a cut-off when needed and
\be
A_N(t)=\prod_{i=1}^{d^*}\int_{-\infty}^{+\infty} dx_i\, \exp\left(-\frac{Nx_i^2}{4Dt}\right)=
\left(\frac{4\pi Dt}{N}\right)^{d^*/2}\,.
\label{ANt}
\ee
Collecting these results leads to
\be
E_N(t)\approx 1+\frac{a^{Nd-d^*}}{\tau(4\pi D)^{(Nd-d^*)/2}N^{d^*/2}}\,\int_{\t0}^t dt'(t')^{(d^*-Nd)/2}\,.
\label{ENt4}
\ee
Let $B_N(t)$ denote the time integral in~\eref{ENt4}. It is given by
\bea
B_N(t)&=\frac{t^{1+(d^*-Nd)/2}-\t0^{1+(d^*-Nd)/2}}{1+(d^*-Nd)/2}\,,\qquad &d\neq d_c(N,d^*)=\frac{d^*+2}{N}\,,
\nonumber\\
B_N(t)&=\ln\left(\frac{t}{\t0}\right)\,,\qquad &d=d_c\,.
\label{BNt}
\eea
Thus the long-time behaviour of the integral changes 
at the critical dimension $d_c$. It is governed by the constant cut-off when $d>d_c$ and by the time-dependent upper limit otherwise. The asymptotic behaviour of the mean number of encounters follows with
\bea
E_N(t)&\sim t^{1+(d^*-Nd)/2}\,,\qquad &0<d<d_c(N,d^*)=\frac{d^*+2}{N}\,,\nonumber\\
&\sim\ln(t/t_0)\,,\qquad &d=d_c\,,\nonumber\\
&\sim{\rm const.}\,,\qquad &d>d_c\,.
\label{ENt5}
\eea
When $d^*=d$ the critical dimension, which is the solution of $d_c(N,d_c)=d_c$, is given by
\be
d_c(N)=\frac{2}{N-1}\,,\qquad d^*=d\,.
\label{dc}
\ee
Actually \eref{ENt4} remains valid for $d=d^*=0$ where $E_N(t)\approx t/\tau$. Then the walkers cannot move and each time step gives an encounter. 

\begin{table}\fl
\caption{Behaviour of the mean cumulated number of encounters $E_N(t)$ in $d=1$. \label{t1}}
\begin{indented}\item[]
\begin{tabular}{@{}cccccc}
\br
$d=1$&$N$ &$1$&$2$&$3$&$4$\\
\mr
&$d^*=0$&$\sim \sqrt{t}$&$\sim\ln(t/t_0)$&$\alpha^1_{03}-\beta^1_{03}\, t^{-1/2}$&$\alpha^1_{04}-\beta^1_{04}\, t^{-1}$ \\ \ms
&$d^*=1$&$t+1$&$\sim \sqrt{t}$&$\sim\ln(t/t_0)$&$\alpha^1_{14}-\beta^1_{14}\, t^{-1/2}$ \\ \ms
\br
\end{tabular}
\end{indented}
\end{table}

\begin{table}\fl
\caption{Behaviour of the mean cumulated number of encounters $E_N(t)$ in $d=2$. \label{t2}}
\begin{indented}\item[]
\begin{tabular}{@{}ccccc}
\br
$d=2$&$N$ &$1$&$2$&$3$\\
\mr
&$d^*=0$&$\sim\ln(t/t_0)$&$\alpha^2_{02}-\beta^2_{02}\, t^{-1}$&$\alpha^2_{03}-\beta^2_{03}\, t^{-2}$\\ \ms
&$d^*=1$&$\sim \sqrt{t}$&$\alpha^2_{12}-\beta^2_{12}\, t^{-1/2}$&$\alpha^2_{13}-\beta^2_{13}\, t^{-3/2}$\\ \ms
&$d^*=2$&$t+1$&$\sim\ln(t/t_0)$&$\alpha^2_{23}-\beta^2_{23}\, t^{-1}$\\ \ms
\br
\end{tabular}
\end{indented}
\end{table}

The behaviour of $E_N(t)$ in one and two dimensions, as obtained in~(\ref{ENt4}--\ref{BNt}), is shown in tables~\ref{t1}--\ref{t2} for different values of $N$ and~$d^*$. The time-independent coefficients are written as $\alpha^d_{d^*N}$ and~$\beta^d_{d^*N}$. When $N=1$, $E_N(t)$ gives 
the mean number of times a single walker visits the subspace of dimension $d^*$ up to time $t$.  When $d^*=d$ the encounters are counted without restriction on their location.
In particular, with $d^*=d$ and $N=1$, the walker has an encounter with himself at each time step, hence $E_N(t)=t+1$.

\section{Monte Carlo simulations}
\begin{figure}[t]
\vspace*{5mm}
  \[
  \includegraphics[width=0.6\textwidth]{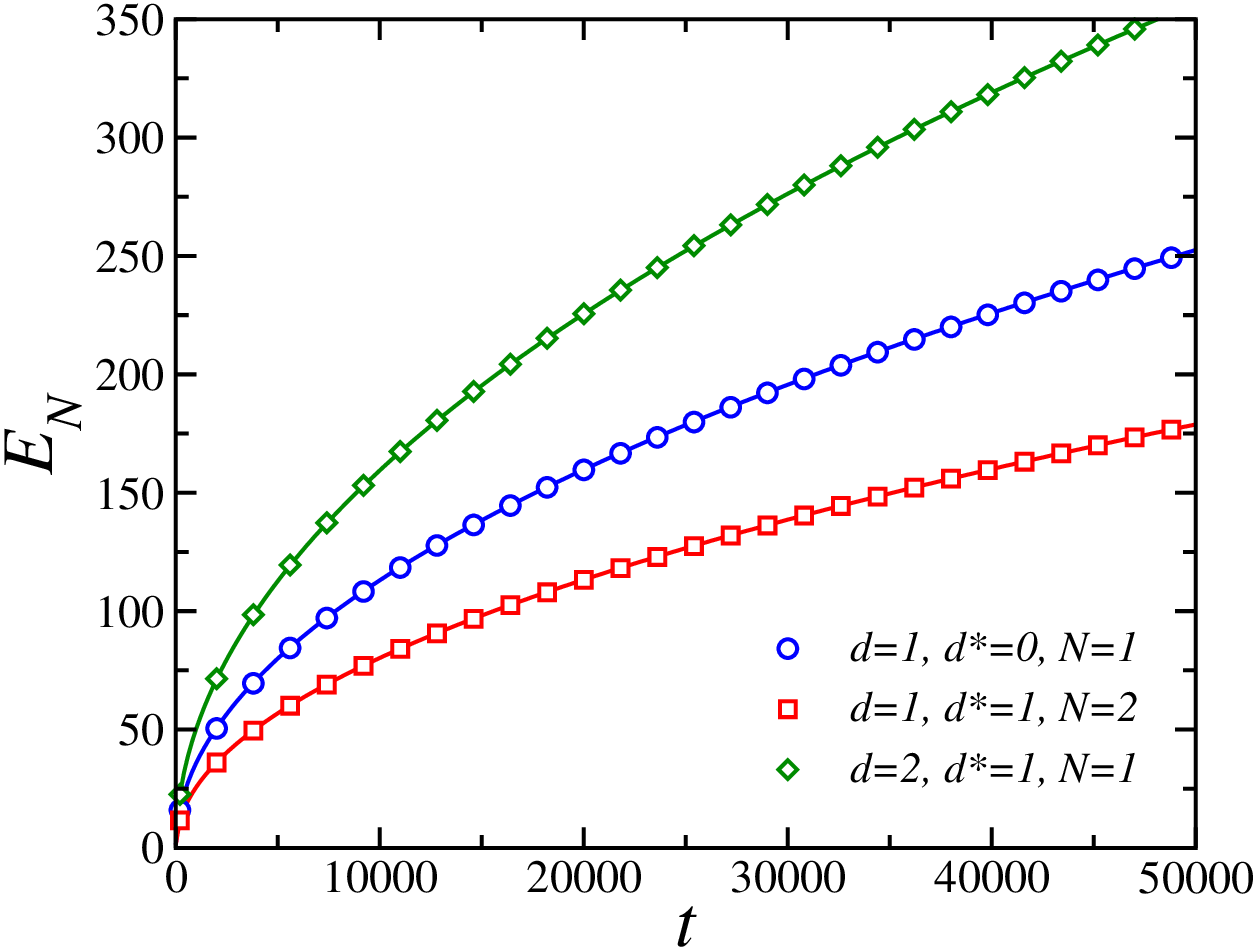}
  \]
\vspace*{-5mm}
  \caption{(Colour online) Comparison between the analytical results in the continuum approximation (lines) and the Monte Carlo data (symbols) for the mean cumulated number of encounters $E_N(t)$ on $\ld*$ of $N$ random walkers starting from the origin of $\ld$ at $t=0$. The values of $N$ and $d^*$ are such that $d$ is below the critical dimension $d_c$ and $E_N$ grows as $\sqrt{t}$. When $N=1$, $E_N$ gives the mean number of visits of the sub-lattice $\ld*$.}
   \label{fig1_enc}
\vspace*{5mm}
\end{figure}

\begin{figure} [b]
\vspace*{5mm}
  \[
  \includegraphics[width=0.6\textwidth]{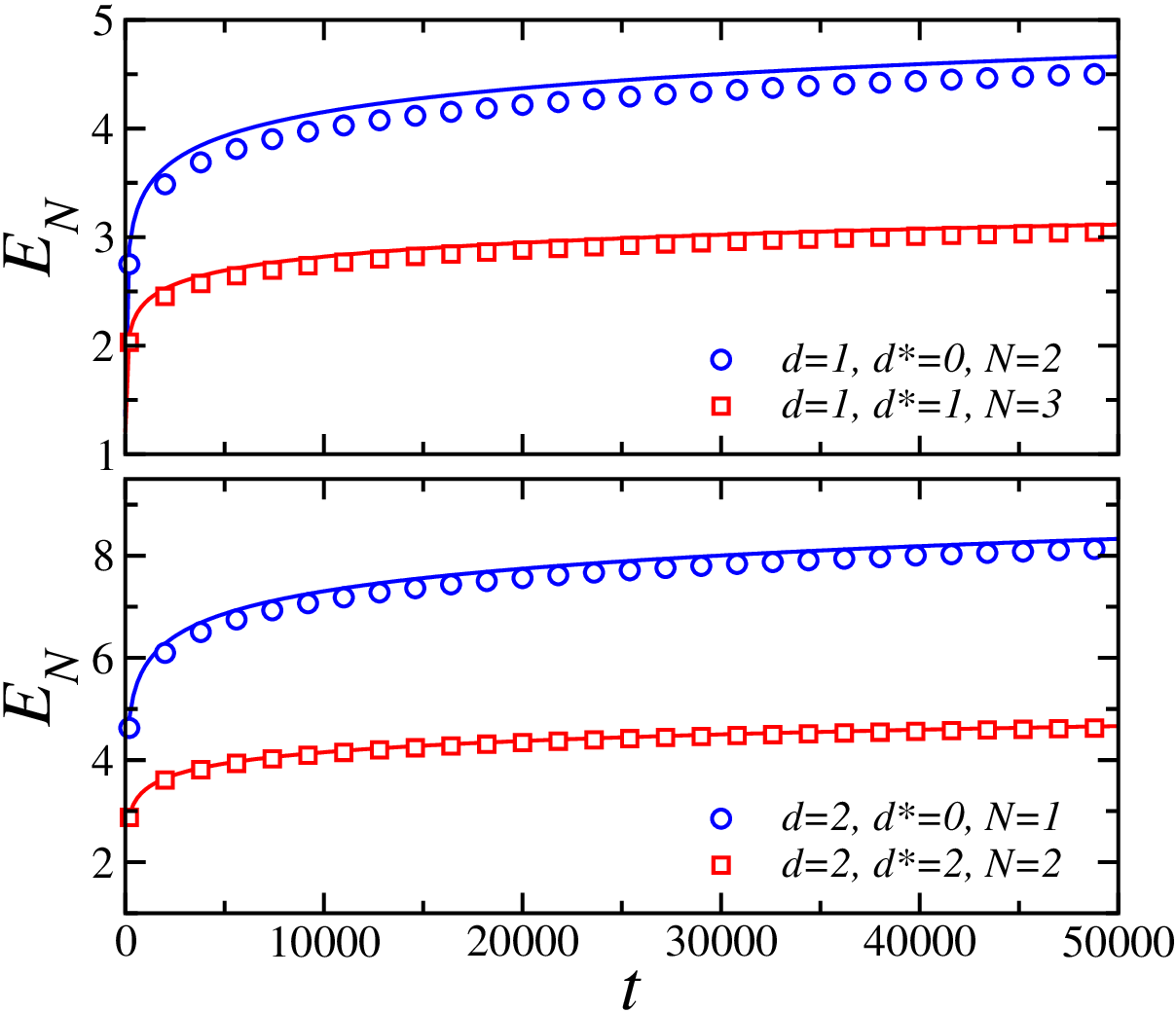}
  \]
\vspace*{-5mm}
  \caption{(Colour online) As in figure \protect\ref{fig1_enc} but the values of $N$ and $d^*$ are such that $d=d_c$. The growth of the mean cumulated number of encounters $E_N$ is logarithmic in $t$.}
  \label{fig2_enc}  \vskip .5cm
\end{figure}

\begin{figure} [t]
\vspace*{12mm}
  \[
  \includegraphics[width=0.7\textwidth]{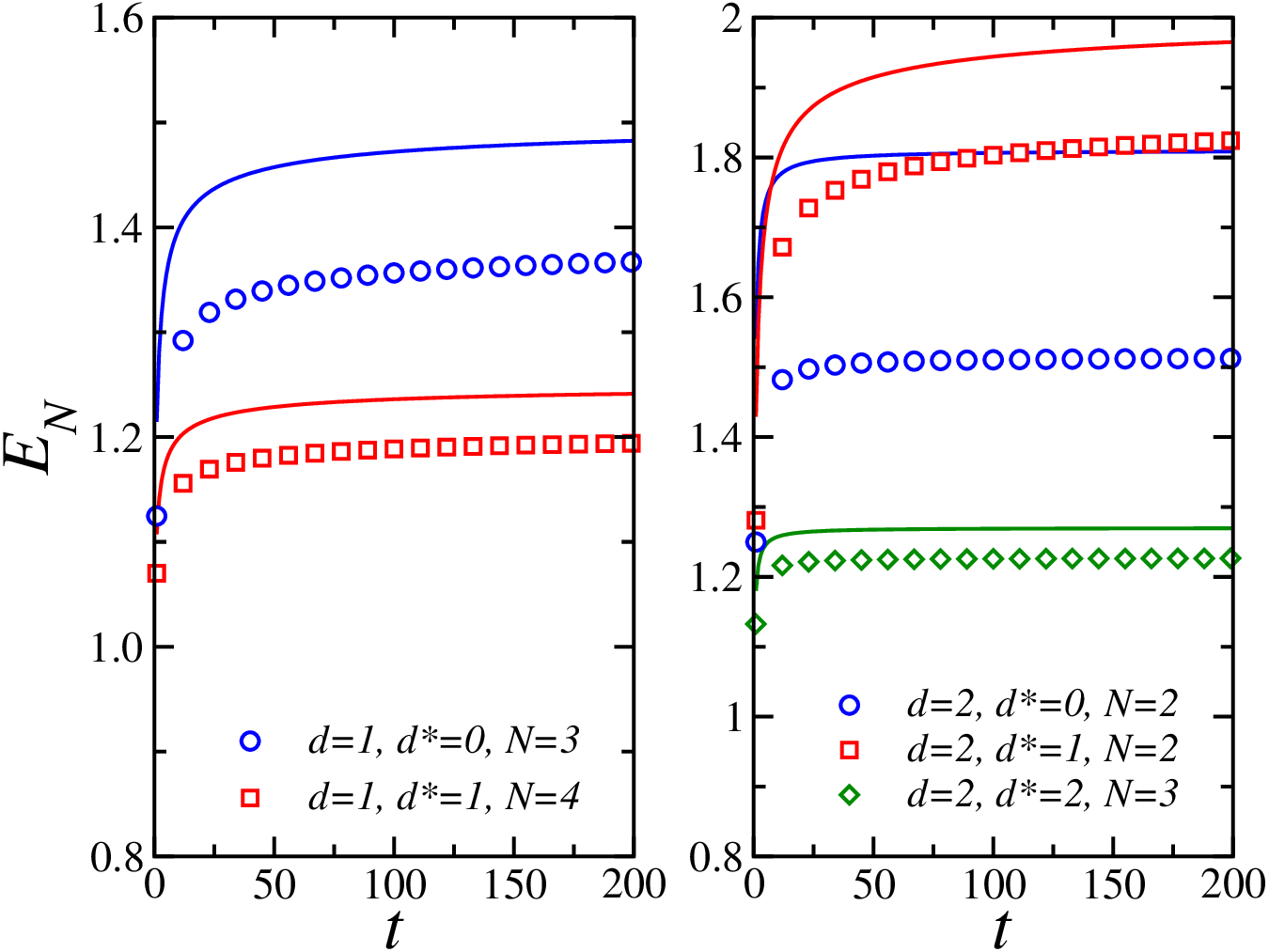}
  \]
\vspace*{-5mm}
  \caption{(Colour online) As in figure \protect\ref{fig1_enc} for values of $N$ and $d^*$ such that $d>d_c$. The mean cumulated number of encounters $E_N$ tends to a small constant value which is approached according to a power law, $t^{-1/2}$ for $d=1$ and for $d=2$ (squares), $t^{-1}$ for $d=2$ (circles and diamonds).}
  \label{fig3_enc} 
\end{figure}

\begin{figure} [b]
\vspace*{0mm}
  \[
  \includegraphics[width=0.7\textwidth]{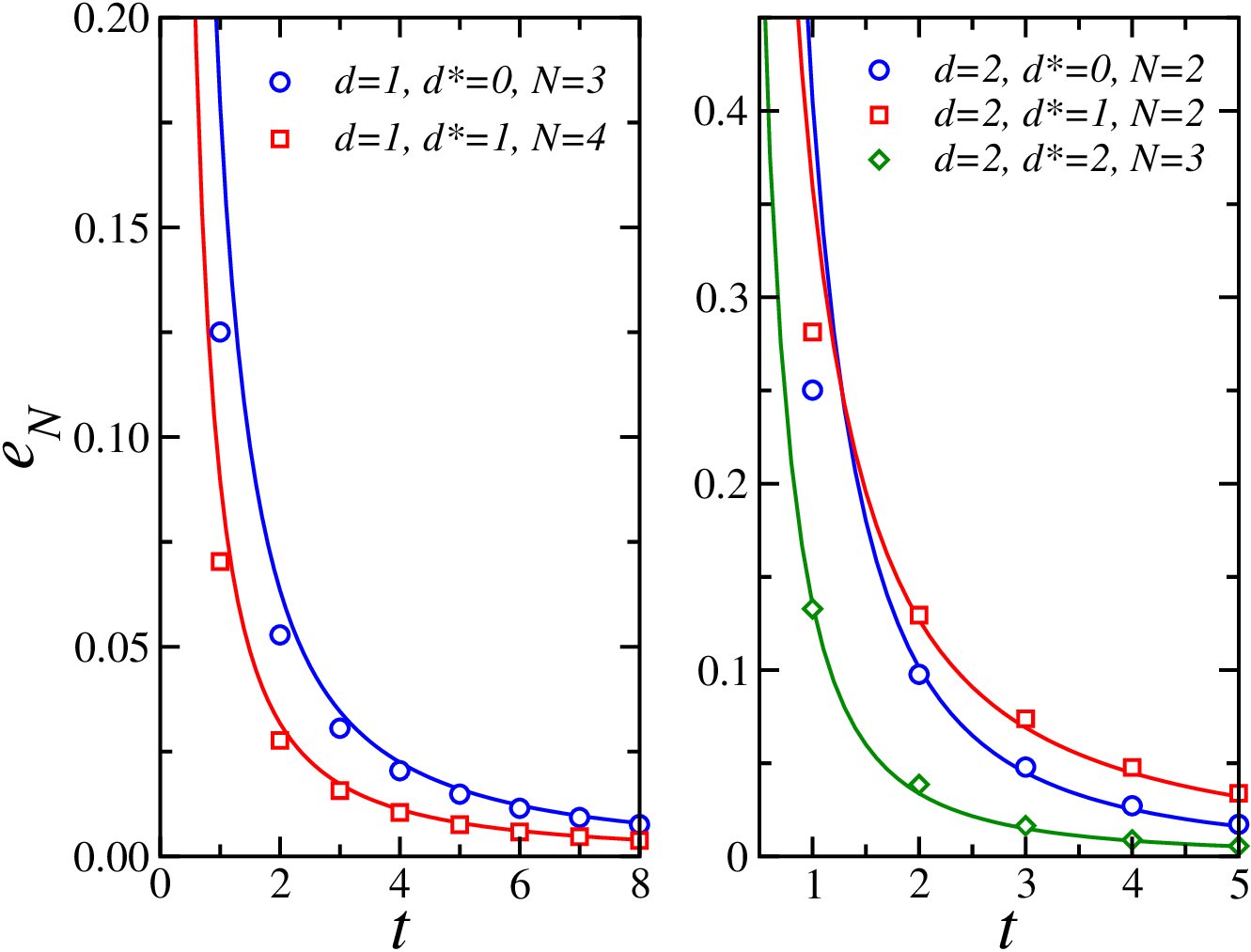}
  \]
\vspace*{-5mm}
  \caption{(Colour online) Mean number of encounters $e_N(t)$ on $\ld*$ of $N$ random walkers starting from the origin of $\ld$ at $t=0$. The results obtained in the continuum approximation (lines) quickly converge to the Monte Carlo data (symbols).}
  \label{fig4_enc}  
\end{figure}

The mean number of encounters of discrete random walkers has been studied numerically via Monte Carlo simulations for different values of $N$, $d$ and $d^*$ in the different regimes around $d_c$, for $d=1$ and $2$. 

The hyper-cubic lattice can be decomposed in two sub-lattices with $\sum_{i=1}^d x_i$ even on one sub-lattice and odd on the other. When the random steps are between nearest neighbour sites only, the random walkers starting from the origin at $t=0$ will stay on the even (odd) sub-lattice when $t$ is even (odd). Due to this parity alternation the probability of encounter is artificially multiplied by a factor of $2$ with respect to continuum results. 

In order to avoid this effect, we break the parity alternation by allowing the walkers to wait on the same lattice site with probability $1/2$ or to jump with probability $1/(4d)$ to one of the nearest neighbour sites. Thus we use the following single step probability density in $d$ dimensions
\be
p(\vec{r})=\frac{1}{2}\,\delta(\vec{r})+\frac{1}{4d}\sum_{i=1}^d\left[\delta(\vec{r}+a\,\vec{u}_i)+\delta(\vec{r}-a\,\vec{u}_i)\right]\,.
\label{stepproba}
\ee
With $a=\tau=1$ the diffusion coefficient is
\be
D=\frac{\langle r^2\rangle}{2d\tau}=\frac{1}{4d}\,.
\label{difcoef}
\ee
The mean number of encounters was evaluated by averaging over $10^6$ samples so that the statistical errors are negligible. 

The Monte Carlo data for $E_N(t)$ are compared to the analytical results of~\eref{ENt4} and~\eref{BNt} 
for values of $d^*$ and $N$ leading to the different regimes: power law growth below $d_c$ 
in figure~\ref{fig1_enc}, logarithmic growth at $d_c$ in figure~\ref{fig2_enc} and power law approach to saturation
above $d_c$ in figure~\ref{fig3_enc}. 

A systematic deviation between the analytical results (lines) and the Monte Carlo data (symbols) becomes apparent
at, and mainly above $d_c$ where the difference is no longer negligible compared to the values of $E_N$.  
It is clearly due to the failure of the continuum approximation at short time. In order to put it in evidence, the mean number of encounters {\it at} time $t$, $e_n(t)$, which is given by $dE_N/dt$ in the continuum approximation, is shown in figure~\ref{fig4_enc} for values of $d^*$ and $N$ leading to the saturation regime.  The continuum results converge to the Monte Carlo data after only a few time steps. 

\section{Intersection of strongly anisotropic fractals} 

The random walk, when considered in space-time, is a strongly anisotropic fractal growing as $t$ in the time direction and as $\sqrt{t}$ in the $d$ transverse spatial directions. Thus counting the encounters of random walkers is a problem of intersection in space-time of strongly anisotropic fractals.

Let us consider a strongly anisotropic fractal in $d+1$ dimensions. It has a fractal dimension $d_f$ and a typical size $R$ in the $d$ transverse directions
while the corresponding values are $d_{f\parallel}$ and $R_\parallel$ in the longitudinal (temporal) direction. The mass of the fractal is
\be
M\sim R^{d_f}\sim R_\parallel^{d_{f\parallel}}\,,\qquad R_\parallel=R^z\,,\qquad d_{f\parallel}=\frac{d_f}{z}\,,
\label{mr}
\ee
its volume is 
\be
V\sim R^d R_\parallel\sim R^{d+z}\,,
\label{vr}
\ee
so that the average density scales as
\be
\langle\rho\rangle=\frac{M}{V}\sim R^{d_f-d-z}\sim R^{-\overline{d_f}}\,,
\label{rho}
\ee
which defines the fractal co-dimension for a strongly anisotropic object
\be
\overline{d_f}=d+z-d_f\geq0\,.
\label{odf}
\ee
Consider now $N$ statistically independent fractals with a common origin and the same values of $d_f$ and $z$. The average density of their intersection, $\langle{\rho_N}\rangle$, will scale as ${\langle\rho\rangle}^N$ so that
\be
\langle{\rho_N}\rangle\sim\langle{\rho}\rangle^N\sim R^{-N\overline{d_f}}\sim R^{-\overline{d_f(N)}}\,,
\label{rhoN}
\ee
where $\overline{d_f(N)}$ is the fractal co-dimension of the intersection of the $N$ fractals. Thus the fractal dimension of the intersection is given by
\be
d_f(N)=d+z-\overline{d_f(N)}\,.
\label{dfN}
\ee
Since the mass of the fractal intersection cannot decay with $R$, $\overline{d_f(N)}$ is bounded from above by $d+z$ so that:
\be
\overline{d_f(N)}=\min [d+z,N\overline{d_f}]\geq0\,.
\label{odfN}
\ee
The co-dimension of the fractal intersection is given by the sum of the co-dimensions of the fractals involved, when this sum is smaller than $d+z$.

For $N$ isotropic fractals with dimension $d_f$ in $D=d+1$ dimensions, $z=1$, $\overline{d_f}=D-d_f$ and, according to~\eref{odfN}, the co-dimension of the fractal intersection is $\overline{d_f(N)}=\min [D,N\overline{d_f}]\geq0$, in agreement with \eref{dfiso} for the Euclidean dimension $D$.

We are actually interested in that part of the fractal intersection which belongs to a subspace with 
dimension $d^*+1$ in space-time, i. e., its intersection with this subspace. According to the rule of co-dimension additivity we have to add the co-dimension $d-d^*=d+1-(d^*+1)$ to $N\overline{d_f}$ in~(\eref{odfN}). The total co-dimension is then
\be
\overline{d_f(N,d^*)}=\min [d+z,N\overline{d_f}+d-d^*]\geq0\,,\qquad \overline{d_f(N,d)}=\overline{d_f(N)}
\label{odfNd}
\ee
and the appropriate fractal dimension reads:
\be
d_f(N,d^*)=d+z-\overline{d_f(N,d^*)}\,.
\label{dfNd}
\ee

The upper critical dimension $d_c(N,d^*)$ is reached when the sum of the co-dimensions in $\overline{d_f(N,d^*)}$ is equal to $d+z$, leading to $d_f(N,d^*)=0$. Then we have $d_c+z=N(d_c+z-d_f)+d_c-d*$ so that:
\be
d_c(N,d^*)=d_f-z+\frac{d^*+z}{N}\,.
\label{dcNd}
\ee
When $d^*=d$, according to~\eref{odf} and~\eref{odfN}, we have $d_c+z=N(d_c+z-d_f)$ which gives:
\be
d_c(N)=\frac{z+N(d_f-z)}{N-1}\,.
\label{dcN}
\ee

Below the critical dimension $d_c(N,d^*)$ the fractal growth of the intersection is governed by
\be
M^*(R)\sim R^{d_f(N,d^*)}\sim R^{z+d^*-N(d+z-d_f)}\qquad  (d<d_c)
\label{mr-0}
\ee
according to \eref{dfNd}.

The logarithmic growth at the critical dimension can be obtained 
by integrating the local density of the fractal intersection $\rho_N(r)\sim r^{-\overline{d_f(N)}}$ over 
the volume $V^*(R)\sim R^{d^*}R^\parallel\sim R^{d^*+z}$ so that
\be
M^*(R)\sim\int_a^R \rho_N(r)\,dV^*(R)\sim\int_a^R r^{-\overline{d_f(N)}+d^*+z-1}dr\,.
\label{mr-1}
\ee
Alternatively, one may integrate the fractal density  $\rho^*_N(r)\sim r^{-\overline{d_f(N,d^*)}}$ over the unrestricted volume $V(R)\sim R^{d+z}$ which gives
\be
M^*(R)\sim\int_a^R \rho^*_N(r)\,dV(R)\sim\int_a^R r^{-\overline{d_f(N,d^*)}+d+z-1}dr\,.
\label{mr-2}
\ee
The two integrals lead to the same result when $d\leq d_c$ since $\overline{d_f(N)}-d^*= \overline{d_f(N,d^*)}-d$ according to~\eref{odfN} and~\eref{odfNd}. At the critical dimension $\overline{d_f(N,d^*)}=d+z$ and
\be
M^*(R)\sim\int_a^R r^{-1}\,dr\sim \ln(R/a)\qquad (d=d_c)
\label{mr-3}
\ee

Above the critical dimension $\overline{d_f(N,d^*)}$ in~\eref{odfNd} sticks to the value 
$d+z$ so that $d_f(N,d^*)=0$ and
\be
M^*(R)\sim {\rm const.}\qquad  (d>d_c)
\label{mr-4}
\ee
asymptotically.

For the random walk we have $R(t)\sim\sqrt{t}$ and $R_\parallel(t)=t$. The mass of the fractal is the number of steps, $M(t)=t$. Thus, according to~\eref{mr}, $d_f=z=2$. With these values of $d_f$ and $z$ the critical dimensions in~\eref{dcNd} for $d^*<d$ and~\eref{dcN} for $d^*=d$ agree with the values given in~\eref{ENt5} and~\eref{dc}. 

The mean cumulated number of encounters at time $t$ of the $N$ random walkers in $\ld*$ is given by the mass of the fractal intersection $M^*(R)$ with $R\sim\sqrt{t}$. Inserting this value of $R(t)$ in~\eref{mr-0},~\eref{mr-3} and~\eref{mr-4} one recovers the asymptotic behaviours of $E_N(t)$ in the different regimes (see \eref{ENt5}).

\section{Concluding remarks} 
According to \eref{BNt}, the exponent governing the time dependence of $E_N$ 
involves $N$, $d$ and $d^*$ through the combination $d^*-Nd$. Thus changing $d^*$ by $\Delta d^*$, $N$ by $\Delta N$ and $d$ by $\Delta d$ does not affect the asymptotic behaviour of $E_N(t)$~\cite{amplitude} when
$\Delta d^*=N\Delta d+d\Delta N$. This is the origin of the correspondences appearing in tables \ref{t1} and \ref{t2}.
For example, in $d$ dimensions up to time $t$, the mean number of simultaneous visits of the origin ($d^*=0$) by $N$ random walkers has the same time dependence as the mean number of encounters of $N+1$ random walkers anywhere ($d^*=d$), a well-known fact for $N=1$. 

Another remark is in order concerning the regime $d>d_c$ where $E_N$ is asymptotically constant. This behaviour follows from~\eref{odfNd} where the minimisation leads to $\overline{d_f(N,d^*)}=d+z$ 
so that $d_f(N,d^*)=0$. If the minimisation is not taken into account, the fractal dimension in \eref{dfNd} takes on negative values when $d>d_c$. Actually, this negative fractal dimension governs the approach to saturation in~\eref{BNt}. 

The notion of fractal intersection introduced here for identical strongly anisotropic fractals  can be extended to the case of fractals with different values of the anisotropy exponent $z$ and/or the fractal dimension $d_f$. 

To conclude let us stress that the notion of fractal intersection used here could be efficient in other similar problems. Among these one may mention the asymptotic behaviour of $W_N$ and $E_N$  for L\'evy flights~\cite{berkolaiko96,berkolaiko98}.

\end{document}